\documentclass[prb, twocolumn, 10pt, final, floatfix, aps]{revtex4-1}

\pdfoutput=1
\usepackage{graphicx}
\usepackage{amsmath, amssymb, amsfonts, bm}
\usepackage{latexsym}
\usepackage[colorlinks]{hyperref}
\usepackage[protrusion=true,expansion=true]{microtype}

\newcommand{\ee}{\mathrm{e}}
\newcommand{\ii}{\mathrm{i}}
\newcommand{\D}[1]{\,\mathrm{d}#1}
\newcommand{\Dev}{\xi}
\newcommand{\JumpR}{\chi}

\begin{document}
\title{Motional Averaging in a Superconducting Qubit}
\author{Jian Li$^1$} \author{M.~P. Silveri$^2$} \author{K.~S. Kumar$^1$}
\author{J.-M. Pirkkalainen$^1$} \author{A. Veps\"al\"ainen$^1$}
\author{W.~C. Chien$^{1}$} \author{J. Tuorila$^2$}
\author{M.~A. Sillanp\"a\"a$^1$} \author{P.~J. Hakonen$^1$}
\author{E.~V. Thuneberg$^2$} \author{G.~S. Paraoanu$^1$}
\affiliation{$^1$O. V. Lounasmaa Laboratory, Aalto University School of Science, P.O. Box 15100,
  FI-00076 AALTO, Finland \\
$^2$Department of Physics, University of Oulu, P.O. Box 3000, FI-90014, Finland }
\date{\today}
\collaboration{Published in Nature Communications \textbf{4}, 1420 (2013). \href{http://dx.doi.org/10.1038/ncomms2383}{http://dx.doi.org/10.1038/ncomms2383}}
\begin{abstract}
{\bf Superconducting circuits with Josephson junctions are promising candidates for developing future quantum technologies. Of particular interest is to use these circuits to study effects that typically occur in complex condensed-matter systems. Here, we employ a superconducting quantum bit (qubit)\----a transmon\----to carry out an analog simulation of motional averaging, a phenomenon initially observed in nuclear magnetic resonance (NMR) spectroscopy. To realize this effect, the flux bias of the transmon is modulated by a controllable pseudo-random telegraph noise, resulting in stochastic jumping of the energy separation between two discrete values. When the jumping is faster than a dynamical threshold set by the frequency displacement of the levels, the two separated spectral lines merge into a single narrow-width, motional-averaged line. With sinusoidal modulation a complex pattern of additional sidebands is observed. We demonstrate experimentally that the modulated system remains quantum coherent, with modified transition frequencies, Rabi couplings, and dephasing rates. These results represent the first steps towards more advanced quantum simulations using artificial atoms. }
\end{abstract}
\maketitle

\begin{figure*}
\includegraphics[width=1.0\linewidth]{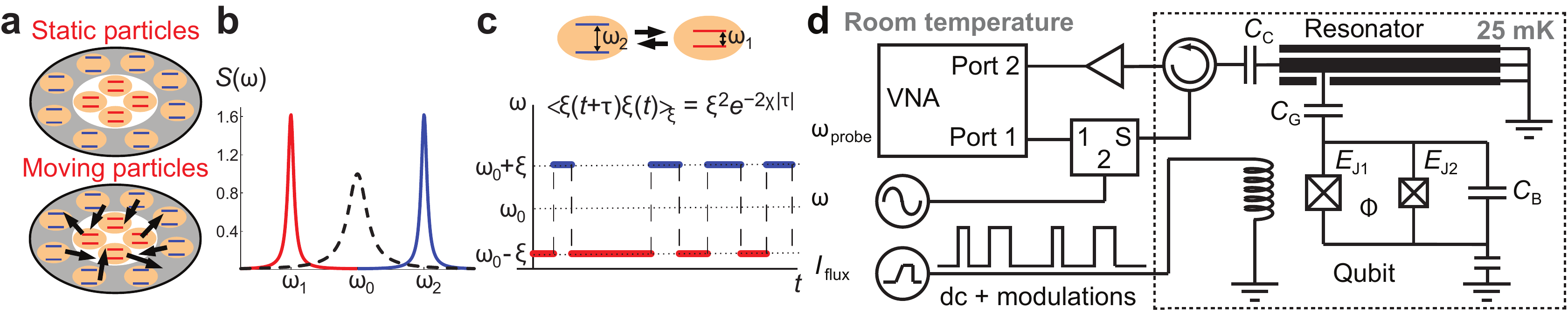}
  \caption{\label{fig:NMR-cavity}{\bf Schematics of motional averaging and simplified experimental set-up.} {\bf a,} Illustration of a single pore filled with a medium consisting of spin-$1/2$ particles that are either static or moving. White and grey areas denote different magnetic environments, producing the energy splittings $\hbar \omega_1$ and $\hbar\omega_2$, respectively (see {\bf b,c}). {\bf b,} Schematic spectrum of the motional averaging. The solid lines denote the spectrum of static particles, while the dashed line corresponds to moving particles. {\bf c,} A sample trajectory of the energy splitting $\hbar[\omega_0+\Dev(t)]$ in equation~\eqref{eq:ham-time-dep}. {\bf d,} Schematic diagram of the sample device and the simplified experimental set-up. Two asymmetric Josephson junctions ($E_{\rm J1}$ and $E_{\rm J2}$) are capacitively shunted ($C_{\rm B}$) and coupled ($C_{\rm G}$) to a resonator. The time-dependent transition frequency of the qubit is controlled with fast flux pulses on top of a dc bias ($I_{\rm flux} = I_{\rm dc} + I_{\rm pulse}$), generated by an arbitrary waveform generator. A vector network analyzer (VNA) is used for probing the resonator at $\omega_{\rm probe}\approx\omega_{\rm r}$. One microwave source ($\omega$) is used for driving the qubit. Additional details are shown in Fig. S1 of Supplementary. }
\end{figure*}

The ability to resolve energy variations $\Delta E$ occurring in a time interval $\Delta t$ is fundamentally limited by the energy-time uncertainty relation $\Delta E \Delta t \gtrsim\hbar$. Consider for example a system of spin-1/2 particles filling a porous material~(Fig.~\ref{fig:NMR-cavity}a) in an external magnetic field; then, if the pore wall is paramagnetic, the particles near it will experience a modified local magnetic field. As a result, static particles will produce two spectral peaks in the NMR spectrum at frequencies $\omega_1$ and $\omega_2$~(Fig.~\ref{fig:NMR-cavity}b). In contrast, particles moving swiftly back and forth between the two regions on timescales $\Delta t$ shorter than $\hbar (\Delta E)^{-1}=(\omega_2 -\omega_1)^{-1}$ are not able to discriminate between the two energy values. Then, as the particles move faster, the outcome in the spectroscopy is not simply a continuous broadening and overlapping of the two peaks. Instead, a new peak emerges at the average frequency $\omega_{0}$ with a width smaller than the energy separation $\Delta E$, denoted as \emph{motional averaging and narrowing}~\cite{Anderson54, Abragam}, respectively. In atomic ensembles and condensed-matter systems, this occurs via fast variations of the electronic state populations, chemical potential, molecular conformation, effective magnetic fields, lattice vibrations, micro-electric fields producing ac-Stark shifts {\it etc.} \cite{Abragam, Anderson54,Jiang08, Kohomoto94, Berthelot06, Sagi10}. Closely related phenomena are the Dyakonov-Perel effect~\cite{DyakonovPerel}, the Dicke line narrowing of Doppler spectra~\cite{Dicke54}, and the quantum Zeno effect~\cite{Milburn88}.

In this paper we report the observation of motional averaging and narrowing in a single artificial atom (a qubit), with a simulated fast-fluctuating environment under direct experimental control. This should be contrasted with the typical situation in condensed-matter systems, where one has a large number of particles and the experimentalist can only indirectly attempt to change the fluctuation rate, typically  by modifying a thermodynamic function of state such as temperature or pressure.

\section*{Results}
Our device is a circuit QED system~\cite{Wallraff04}, consisting of a tunable transmon coupled to a quarter-wavelength coplanar waveguide resonator. The circuit schematic and the simplified experimental set-up are shown in Fig.~\ref{fig:NMR-cavity}d. The applied bias flux $\Phi$ through the qubit loop consists of a constant part $\Phi_{\rm dc}$ and a time-dependent part $\Phi_{\rm ac}(t)$ with an amplitude much smaller than $\Phi_{\rm dc}$.  The Hamiltonian of the transmon qubit~\cite{Koch07} is
 \begin{equation}
 \hat H(t) =\frac\hbar 2 [\omega_0+\Dev(t)]\hat{\sigma}_z , \label{eq:ham-time-dep}
  \end{equation}
where $\omega_0=\omega_{\rm p}[\cos^2(\pi\Phi_{\rm dc}/\Phi_0)+d^2\sin^2(\pi\Phi_{\rm dc}/\Phi_0)]^{\frac1 4} - E_{\rm C}/\hbar$.   Here, $\Phi_0=h/2e$ is the magnetic flux quantum, the Josephson plasma frequency is $\omega_{\rm p}=\sqrt{8E_{\rm C}E_{\rm J}}/\hbar$, $\hat{\sigma}_{x,y,z}$ denote the Pauli matrices, the single-electron charging energy is $E_{\rm C}= e^2/2(C_{\rm G}+C_{\rm B})\approx h\times 0.35$~GHz, $E_{\rm J} = E_{\rm J1} + E_{\rm J2} \approx 24 E_{\rm C}$ is the maximum Josephson energy, and $d = (E_{\rm J1} - E_{\rm J2}) / E_{\rm J}\approx 0.11$ denotes the junction asymmetry. We choose $\Phi_{\rm dc}$ so that $\omega_0 /2\pi = 2.62$~GHz, which is far detuned from the bare resonator frequency $\omega_{\rm r}/2\pi = 3.795$~GHz, allowing the dispersive measurement of the qubit through the resonator by standard homodyne and heterodyne techniques~\cite{Wallraff05}.

The time-dependent part $\hbar\Dev(t)$ of the energy splitting, determined by $\Phi_{\rm ac}(t)$, is controlled by an arbitrary waveform generator via a fast flux line. The random telegraph noise (RTN) is realized by feeding pseudo-random rectangular pulses to the on-chip flux bias coil. Ideally, the dynamics of $\hbar\Dev(t)$ is a stationary, dichotomous Markovian process, characterized by an average jumping rate $\JumpR$ and symmetrical dwellings at frequency values of $\pm\Dev$ (see Fig.~\ref{fig:NMR-cavity}c). The number of jumps is a Poisson process with the probability  $P_n(t)=(\JumpR t)^n e^{-\JumpR t}/n!$ for exactly $n$ jumps within a time interval $t$.  This Poissonian process simulates the temporal variations causing motional averaging in atomic ensembles and condensed-matter systems (Fig.~\ref{fig:NMR-cavity}).

\begin{figure*}
\includegraphics[width=1.0\linewidth]{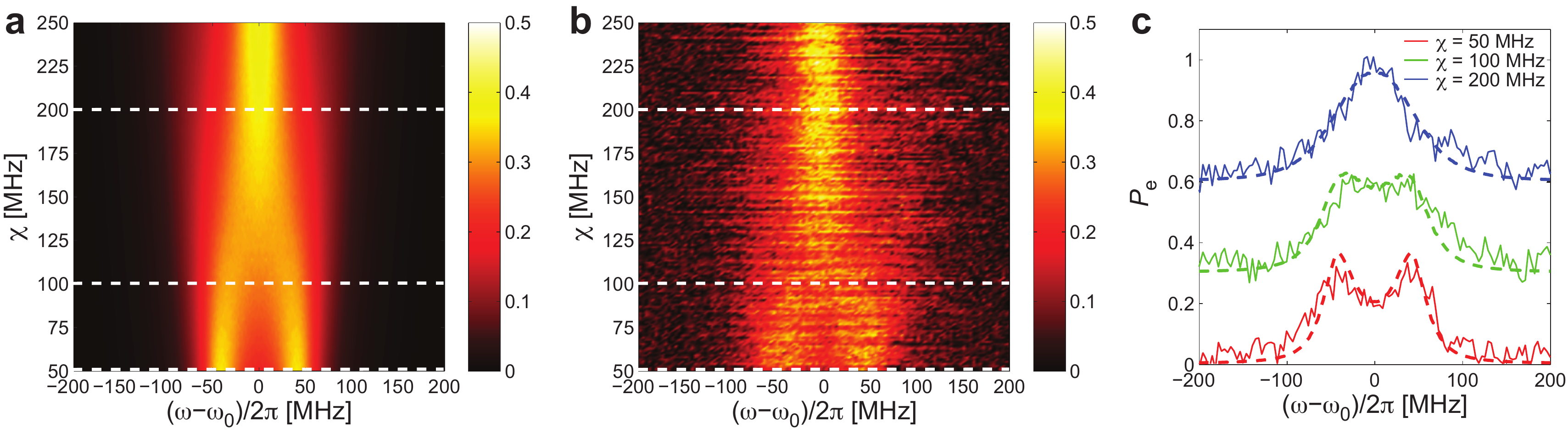}
\caption{\label{fig:mot_aveg}{\bf Motional averaging.} {\bf a,} Numerical simulation of the occupation probability $P_{\rm e}$ plotted in the $\omega-\chi$ plane with parameters $\Dev/2\pi=43$~MHz, $g/2\pi=8$~MHz, $\Gamma_1/2\pi=1$~MHz, and $\Gamma_2/2\pi=3$~MHz. {\bf b,} Measured excited state occupation probability $P_{\rm e}$ under the RTN modulation. {\bf c,} Horizontal cuts of the measured occupation probability in {\bf b}, denoted with white lines,  at three different jumping rates (solid lines). For clarity, the curves are displaced vertically by $0.3$. The dashed lines indicate the corresponding results of the numerical simulation in {\bf a}.  See Fig.~S2 of Supplementary for the zero-detuning  ($\omega=\omega_0$) occupation probability $P_e$ at different $\xi$.}
\end{figure*}

In order to find the effective transition energies and decoherence rates of the modulated system, we calculate the absorption spectrum~\cite{Anderson54,Abragam}. For a qubit subjected to RTN fluctuations as in equation~\eqref{eq:ham-time-dep}, we exploit the quantum regression theorem~\cite{GardinerZoller} to find the spectrum
\begin{equation}
  S(\omega) =\frac{1}{2\pi}\int_{-\infty}^{\infty}\ee^{\ii(\omega-\omega_0)\tau-\Gamma_2|\tau|} \left\langle \ee^{-\ii\int_0^\tau\Dev(t)\D{t}}\right\rangle_\Dev\D{\tau} \label{eq:spect-random},
  \end{equation}
where the expectation value is taken over noise realizations and $\Gamma_2 = \Gamma_\varphi + \Gamma_1/2$ is the decoherence rate in the absence of modulation. For our sample $\Gamma_2 \approx 2\pi \times 3$~MHz (determined in an independent measurement). For the Poisson process $\xi(t)$, Eq.~\eqref{eq:spect-random} gives~\cite{Abragam, Anderson54, Laser84, Mukamel78} (see details in Supplementary)
\begin{equation}
 S(\omega)= \frac{1}{\pi}\frac{2\JumpR\Dev^2+\Gamma_2\left(\bar{\Gamma}_2^2+\bar{\omega}^2+\Dev^2\right)}{(\Dev^2-\bar{\omega}^2+\Gamma_2\bar{\Gamma}_2)^2+4(\Gamma_2+\JumpR)^2\bar{\omega}^2} , \label{eq:spect-mot-aveg-full}
\end{equation}
with $\bar{\Gamma}_2=\Gamma_2+2\JumpR$ and $\bar{\omega}=\omega-\omega_0$. In our measurement, we have recorded the population of the qubit's excited state, as shown in Fig.~\ref{fig:mot_aveg}. The qubit is excited with a transverse drive
\begin{equation}
\hat{H}_{\rm drive}(t)=\hbar g\cos(\omega t)\hat{\sigma}_x. \label{eq:drive}
\end{equation}
The simulation of the occupation probability presented in Fig.~\ref{fig:mot_aveg}a (see Methods and Supplementary) includes also the effect of power broadening due to strong driving amplitudes ($g>\Gamma_1, \Gamma_2$).

Two intuitively-appealing limits result from Eq.~\eqref{eq:spect-mot-aveg-full} (see also Fig.~\ref{fig:mot_aveg}) at the opposite sides of a dynamical threshold defined by $\chi=\xi$. In the case of slow jumping, $\JumpR\ll\Dev$, and resolvable energy variations, $\Delta E \Delta t \sim \hbar\Dev/\JumpR\gg\hbar$, the qubit absorbs energy at $\omega_0\pm\Dev$ with the total decoherence rate $\Gamma_2'=\Gamma_2+\JumpR$. The correlation time of the displacement process $\hbar\Dev(t)$ is $\tau_r=(2\JumpR)^{-1}$ and the linewidth (full width at half maximum $2\Gamma_2'$) broadens by the amount of the reduced mean life-time of a qubit excitation (exchange broadening~\cite{Anderson54}). In contrast, for fast jumping processes, $\JumpR\gg\Dev$, and when the variations are not resolvable, $\Delta E \Delta t\sim \hbar\Dev/\JumpR\ll\hbar$, the qubit absorbs energy only at the frequency $\omega_0$ with $\Gamma_2'=\Gamma_2+\Dev^2/2\JumpR$. The increase in decoherence rate by $\Dev^2/2\JumpR$ can be related to the excursions of the accumulated phase $u(\tau)=\int_0^\tau \Dev(t)\D{t}$ [cf. equation~\eqref{eq:spect-random}] for single noise realizations, especially to the diffusion coefficient of the process~\cite{Dykman10}. Surprisingly, the averaged spectral line is well-localized around the mean value, since $\Dev^2/2\JumpR \ll 2\Dev$. The effect of motional narrowing can be best seen in Fig.~\ref{fig:mot_aveg}c as a raise in the height of the center peak because  the additional decoherence $\Dev^2/2\JumpR$ decreases with increasing jumping rate~$\JumpR$. When $\JumpR$ is comparable with $\Dev$, there is a cross-over region where absorption is reduced and the peak broadens due to enhanced decoherence (see Fig.~\ref{fig:mot_aveg}). This is important for the improvement of qubit dephasing times since it implies that a longitudinally coupled two-level system (TLS) fluctuator is most poisonous when its internal dynamics occurs approximately at the same frequency as the coupling to the qubit.

A counter-intuitive aspect of motional averaging is that the system is at any time in either one of the states $\pm\hbar\Dev$ and spends no time in-between. Yet spectroscopically it is not seen in the states $\pm\hbar\Dev$; instead it has a clear signature in the middle. Also, the  spectrum~\eqref{eq:spect-mot-aveg-full} decays as $1/\omega^{4}$ far in the wings, showing a non-Lorentzian character that originates from the non-exponential decay of the correlator $\left\langle\hat{\sigma}_-(\tau)\hat{\sigma}_+(0)\right\rangle$~\cite{Laser84}. This is because the frequent $\hat{\sigma}_z$-changes have a similar effect to a continuous measurement of the system, slowing its dynamics in analogy with the quantum Zeno effect~\cite{Milburn88, Sagi11}.

\begin{figure*}
\includegraphics[width=1.0\linewidth]{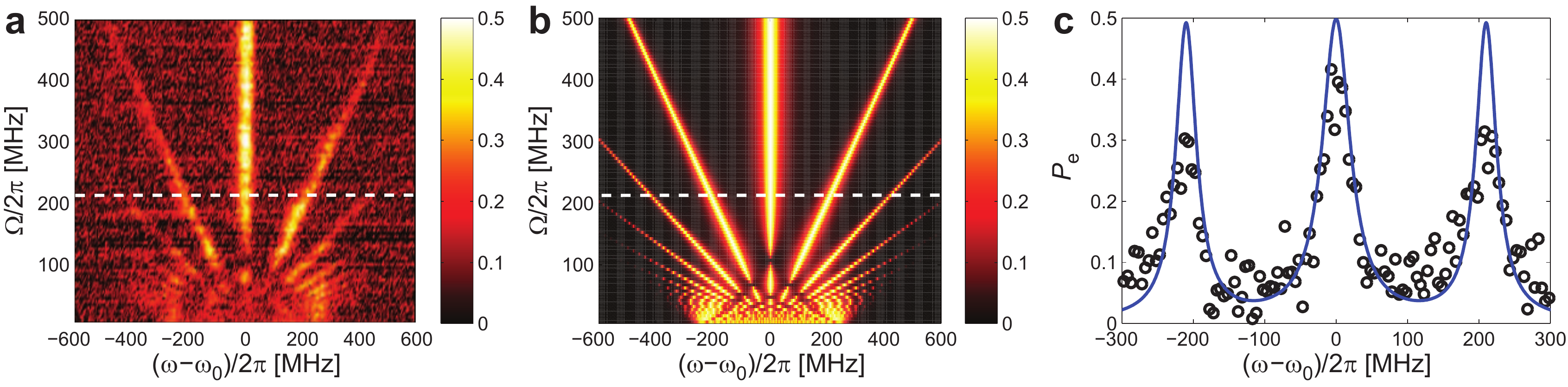}
\caption{\label{fig:sinusoidal} {\bf Occupation probabilities of the excited state of the qubit under sinusoidal modulations.} {\bf a,} Measured excited state occupation probability $P_{\rm e}$ with amplitude $\delta/2\pi=250$~MHz. {\bf b,} Simulation with $\Gamma_1 / 2\pi = 1$~MHz, $\Gamma_2 / 2\pi = 3$~MHz, and $g  / 2\pi = 20$~MHz. {\bf c,} The cut along the dashed line in {\bf a,b} at $\Omega/2\pi=210$~MHz. The (black) circles are the experimental data. The (blue) line denotes the analytical occupation probability obtained from equation~\eqref{eq:sin} by taking $k=-1,0,1$, which overlaps perfectly with the simulation in the parameter regime, $\Omega\gg g>\Gamma_2$. For $k=0$ ($0$-band), the linewidth is narrower than the bare linewidth without modulation (see Fig. S4 of Supplementary).}
\end{figure*}

To further explore these effects, we have used sinusoidal waves to modulate the qubit energy splitting: $\hbar[\omega_0+\delta\cos(\Omega t)]$. In Fig.~\ref{fig:sinusoidal}a,b, the experimental data and the corresponding numerical simulation are shown. Resolved sidebands appear at $\omega=\omega_0\pm k \Omega$ ($k=0, 1,2,\ldots$), and are amplitude-modulated with Bessel functions $J_k(\delta/\Omega)$, where $\delta$ and $\Omega$ denote the modulation amplitude and frequency, respectively.  The Hamiltonian $\hat{H}+\hat{H}_{\rm drive}$  [equations~\eqref{eq:ham-time-dep} and~\eqref{eq:drive}] can be transformed to a non-uniformly rotating frame~\cite{Tuorila10, Oliver05} (see Supplementary). The theoretical prediction for the steady-state occupation probability on the resolved sidebands ($\Omega>g>\Gamma_2$) is
\begin{equation}
P_{\rm e}=\sum_{k=-\infty}^{\infty} \frac{\frac{\Gamma_2}{2\Gamma_1} \left[g J_k\left(\frac{\delta}{\Omega}\right)\right]^2}{\Gamma^2_2+\left(\omega_0+k\Omega-\omega\right)^2+\frac{\Gamma_2}{\Gamma_1}\left[g J_k\left(\frac{\delta}{\Omega}\right)\right]^2},
\label{eq:sin}
\end{equation}
which is compared to experimental data in Fig.~\ref{fig:sinusoidal}c. These population oscillations can be alternatively understood as a photon-assisted version of the standard Landau-Zener-St\"uckelberg interference ~\cite{Noel98, Nori10} (see Fig. S5 of Supplementary). Also, when the qubit drive is tuned close to the resonator frequency $\omega\approx\omega_r$ and when the driving amplitude $g$ is large and equal to the frequency of the modulation $g=\Omega$, the system can be seen as a realization of a quantum simulator of the ultrastrong coupling regime~\cite{Ballester11}. In our setup, the simulated coupling rate is estimated to reach over $10$~\% of the effective resonator frequency (see Supplementary), comparable to earlier reports~\cite{Niemczyk10, FornDiaz10} where the ultrastrong coupling was obtained by sample design. The sinusoidal modulation allows a different perspective on motional averaging. The RTN noise comprises many different modulation frequencies $\Omega$ (Fig.~\ref{fig:NMR-cavity}c). Each frequency creates different sidebands ($k=1,2,3,\ldots$), which overlap and average away. At large $\Omega$'s only the central band $k=0$ survives because the only nonvanishing Bessel function of zero argument is $J_{0}(0)=1$.

\begin{figure*}
\includegraphics[width=1.0\linewidth]{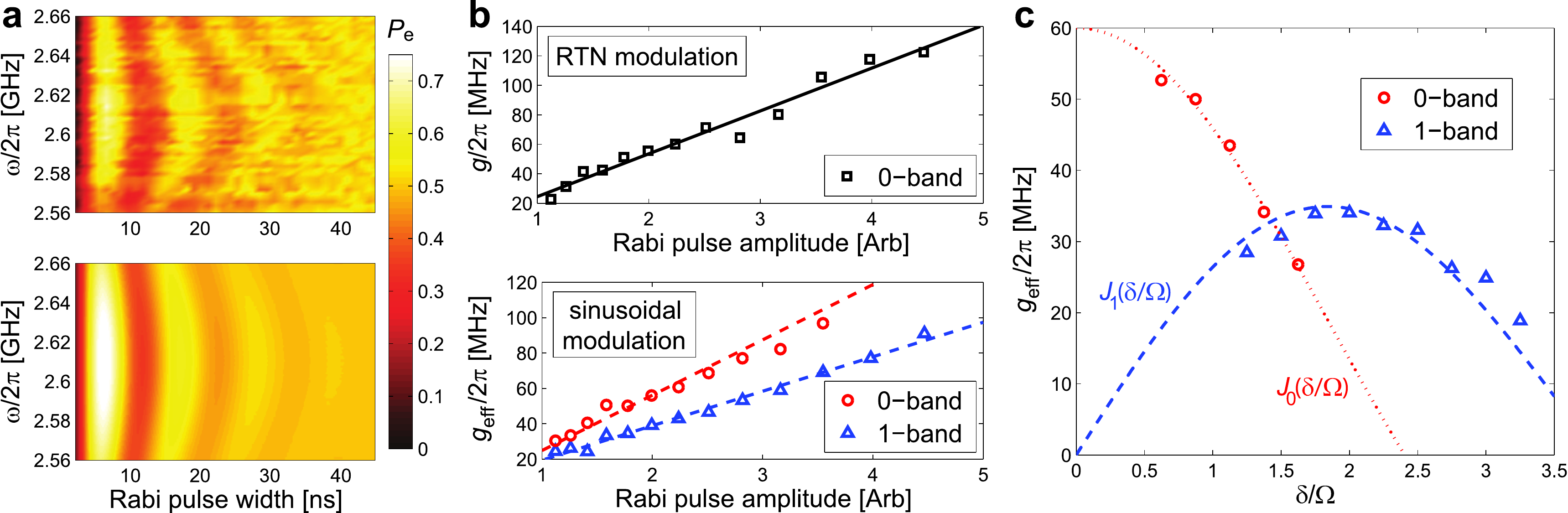}
\caption{\label{fig:rabi}{\bf Rabi oscillations with modulations}. {\bf a,} Rabi oscillations versus driving frequency with RTN modulation: measurement and numerical simulation with $\omega_0/2\pi=2.610$ GHz, $\Dev/2\pi\approx 95$~MHz, $\JumpR\approx 600$~MHz, $g/2\pi = 91$~MHz, $\Gamma_1/2\pi = 1$~MHz, and $\Gamma_2/2\pi = 3$~MHz. The corresponding Rabi oscillations in the case of sinusoidal modulation are shown in Fig. S3 of Supplementary. {\bf b,} The measured Rabi frequency of the random and sinusoidal modulation $g$ and $g_{\rm eff}$, respectively, as a function of Rabi-pulse amplitude (same scaling in the both panels). The solid and dashed lines are fits to the linear portion of the data.  {\bf c,} Rabi frequencies versus amplitude of the sinusoidal modulation. The horizontal axis is the dimensionless scaled modulation amplitude $\delta/\Omega$, which is also the argument for the effective Rabi frequency $g_{\rm eff}^{(k)}=g |J_k(\delta/\Omega)|$ plotted with $g/2\pi=60$ MHz and color coding for $k$: $0$ (red dotted) and $1$ (blue dashed). The (red) circles and (blue) triangles indicate the measured Rabi frequencies at the centre of $0$-band and $1$-band, respectively.}
\end{figure*}

 Finally, we show that it is possible to use the modulated system as a new, photon-dressed qubit. Indeed, we can drive Rabi oscillations on the central band and on the resolved sidebands (see Fig.~\ref{fig:rabi} and Fig.~S3 in Supplementary). For fast RTN modulation, $\JumpR\gg\Dev$, and under the strong drive in equation~\eqref{eq:drive}, we construct (see Supplementary) a master equation~\cite{Mukamel78} describing the Rabi-oscillation with the detuned Rabi frequency $g_{\rm d} = \sqrt{(\omega - \omega_0)^2 + g^2}$ and with the total decoherence rate $\Gamma_2'=\Gamma_2+\xi^2/2\chi$. In the case of sinusoidal modulation, the Rabi frequencies on the sidebands are obtained as $g_{\rm eff}^{(k)}=g |J_k(\delta/\Omega)|$.

\section*{Discussion}
To conclude, we have demonstrated the first realization of motional averaging and narrowing in a single-particle system, {\it i.e.}, a superconducting transmon whose energy splitting was modulated between two predetermined values while maintaining random dwell times in these two states. We have also shown that the system preserves quantum coherence under modulation, with reduced power broadening on the central band. Our work could open up several research directions. For example, a key limitation in solid-state based quantum processing of information is the decoherence due to the fluctuating TLSs  in the dielectric layers fabricated on-chip~\cite{Shnirman05,Martinis05,Yoshihara06}. 
We anticipate, resting on the motional narrowing phenomenon, that the dephasing times of the existing superconducting qubits can be dramatically improved if one can accelerate the dynamics of the longitudinally-coupled TLSs. The quantum coherence of the resolved sidebands and of the central band suggest that low-frequency modulation offers an additional tool for implementing quantum gates, in analogy with the use of vibrational sidebands in the ion trap computers~\cite{SoderbergMonroe10}. 
Also, if the two tones used to drive  and modulate the qubit satisfy certain conditions, the system can be mapped into an effective qubit-harmonic oscillator system with ultrastrong coupling, opening this regime for experimental investigations.
Finally, the recent progress in the field of nanomechanical oscillators makes possible the study of frequency jumps in 
nanomechanical resonators~\cite{Dykman10}, which is predicted to be accompanied by squeezing~\cite{Graham87}. 
Our work paves the way for further simulations of quantum coherence phenomena using superconducting quantum circuits~\cite{Houck12, Friedenauer08}.

\section*{Methods}
Numerical simulations resulting in the occupation probability of Figs.~\ref{fig:mot_aveg}a,c,~\ref{fig:sinusoidal}b, \ref{fig:rabi}a, and S2 and S5c in Supplementary are done by solving the master equation
\begin{align}
\frac{\D{\hat{\rho}(t)}}{\D{t}}=-&\frac{\ii}{\hbar}\left[\hat{H}(t)+\hat{H}_{\rm drive}(t),\hat{\rho}(t)\right]-\frac{\Gamma_\varphi}{4}[\hat{\sigma}_z,[\hat{\sigma}_z,\hat{\rho}]] \notag\\
&+\frac{\Gamma_1}{2}\left(2\hat{\sigma}_-\hat{\rho}\hat{\sigma}_+-\hat{\sigma}_+\hat{\sigma}_-\hat{\rho}-\hat{\rho}\hat{\sigma}_+\hat{\sigma}_-\right), \notag
\end{align}
in the frame rotating at $\omega$ and applying the rotating wave approximation (the pure dephasing rate $\Gamma_\varphi=\Gamma_2-\Gamma_1/2$). For the random modulation (Fig.~\ref{fig:mot_aveg}a,c and ~\ref{fig:rabi}a), we apply the method of quantum trajectories~\cite{GardinerZoller} and the imperfections of the experimentally realized wave forms [the raising/falling time ( $\sim 2.0$~ns) and the sampling time ($0.5$~ns)] are included in the simulation.


{\bf Acknowledgements}\\
We thank S. Girvin and J. Viljas for discussions.  G.S.P. acknowledges the support from the Academy of Finland, projects 141559, 253094, and 135135. M.P.S. was supported by the Magnus Ehrnrooth Foundation and together with J.-M.P. by the Finnish Academy of Science and Letters (Vilho, Yrj\"o ja Kalle V\"ais\"al\"a Foundation). J.-M.P. acknowledges also the financial support from Emil Aaltonen Foundation and KAUTE Foundation. The contribution of M.A.S. was done under an ERC Starting Grant. J.L. and M.P.S. acknowledge partial support from NGSMP.

{\bf Author Contributions}\\
J.L., K.S.K., and G.S.P. designed and performed the experiment. M.P.S. carried out the theoretical work.  A.V., W.C.C., and M.P.S. contributed to experiments. J.-M.P. fabricated the sample. J.T. and E.V.T. provided theoretical support. J.L. and M.P.S. co-wrote the manuscript in co-operation with all the authors. M.A.S., P.J.H., E.V.T., and G.S.P. provided support and supervised the project.

\end{document}


\title{Motional Averaging in a Superconducting Qubit: Supplementary Information}
\author{Jian Li$^1$} \author{M.~P. Silveri$^2$} \author{K.~S. Kumar$^1$}
\author{J.-M. Pirkkalainen$^1$} \author{A. Veps\"al\"ainen$^1$}
\author{W.~C. Chien$^{1}$} \author{J. Tuorila$^2$}
\author{M.~A. Sillanp\"a\"a$^1$} \author{P.~J. Hakonen$^1$}
\author{E.~V. Thuneberg$^2$} \author{G.~S. Paraoanu$^1$}

\affiliation{$^1$O. V. Lounasmaa Laboratory, Aalto University School of Science, P.O. Box 15100,
  FI-00076 AALTO, Finland\\$^2$Department of Physics, University of Oulu, P.O. Box 3000, FI-90014, Finland}
\date{\today}
\maketitle

 \renewcommand{\theequation}{S\arabic{equation}}
\renewcommand{\thefigure}{S\arabic{figure}}

 \section{Measurement set-up}

\begin{figure}[h!]
  \includegraphics[width=0.90\linewidth]{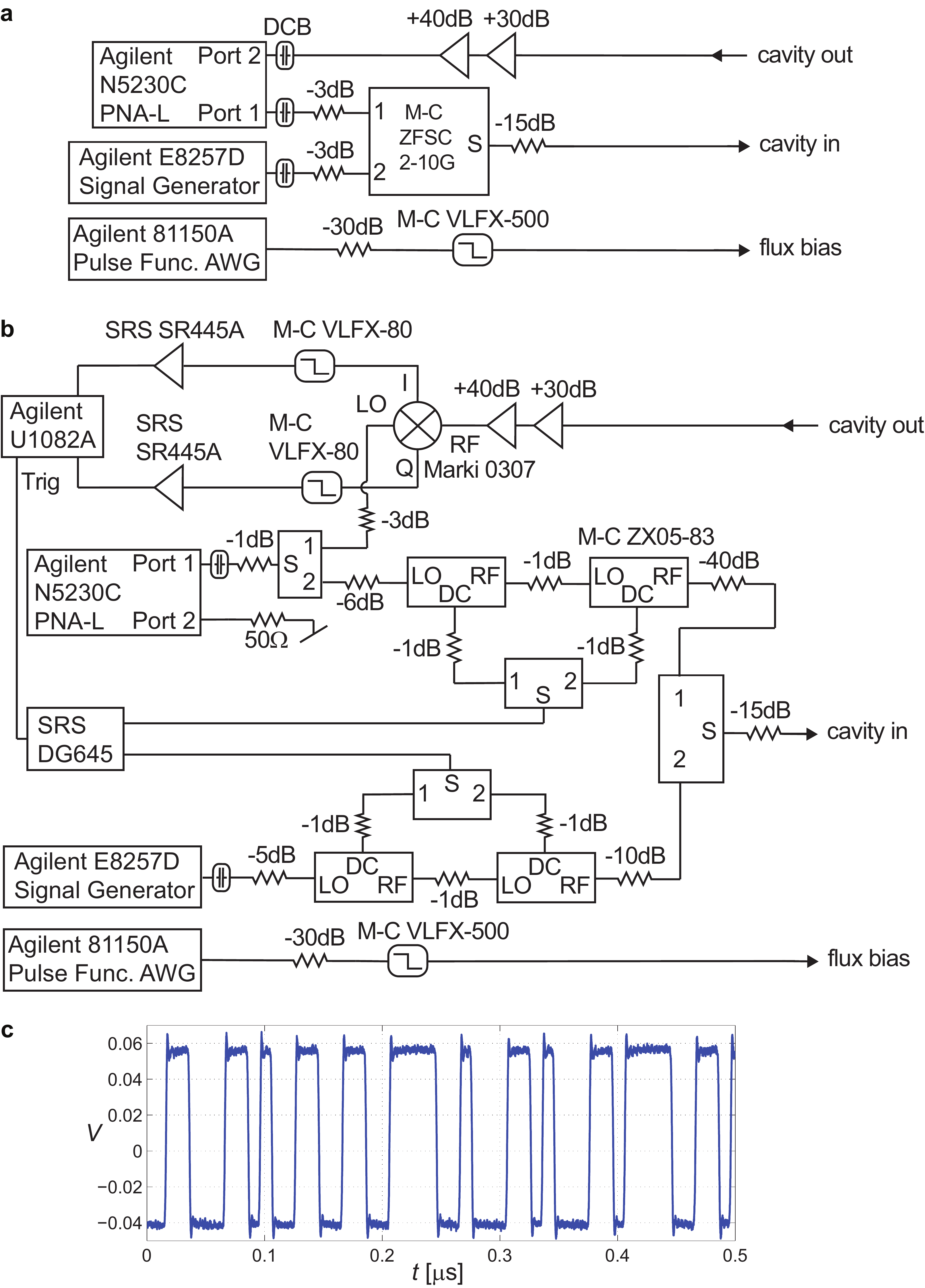}
\caption{\label{room_electronics} {\bf Electronic measurement set-up.} {\bf a,} Detailed diagram of room temperature electronics for qubit spectroscopy. {\bf b,} Detailed diagram of room temperature electronics for Rabi oscillations. {\bf c,} An example of RTN modulation sequence generated by the AWG at a clock frequency $\nu = 10$~KHz, taken by an oscilloscope.}
\end{figure}

The electronic measurement set-up at room temperature is illustrated in Fig.~\ref{room_electronics}. For qubit spectroscopy with RTN modulations, the dc flux bias and RTN modulations are generated by an Agilent 81150A arbitrary waveform generator (AWG). The qubit driving signal from an Agilent E8257D analogue signal generator and the CPW cavity probe signal from an Agilent N5230C PNA-L network analyzer are combined together by a Mini-Circuits ZFSC-2-10G power splitter/combiner and sent to the cavity input line of the dilution refrigerator. The signal from the cavity output line of the dilution refrigerator is amplified and detected by a PNA-L network analyzer. For sinusoidal modulations, the tones are generated by a R\&S SMR27 microwave signal generator and they are added to the dc bias generated by the Agilent 81150A via a bias tee (not shown). Between the radio frequency instruments (analog signal generator and network analyzer) and their corresponding lines, dc blocks (DCBs) are used for breaking possible ground loops. All instruments are synchronized with a SRS FS725 Rubidium frequency standard (not shown in the figure).

For Rabi oscillation measurements, high ON/OFF ratio Rabi pulses are generated by mixing a continuous microwave signal from the Agilent E8257D with rectangular pulses from a SRS DG645 digital delay generator via two identical Mini-Circuits frequency mixers. The Agilent N5230C PNA-L is used as a signal generator. Its output signal is split into two parts. One part is used for generating measurement pulses in a similar fashion to the Rabi pulses; the other part acts as a local oscillator (LO) signal to mix the output signal from the CPW cavity down via a Marki IQ mixer. The IQ data is filtered, amplified by a SRS SR445A pre-amplifier, and digitised by an Agilent U1082A-001 digitiser.

Measurement controlling and data processing are done by MATLAB running on a measurement computer. The communication between the measurement computer and the instruments is realized through IEEE-488 GPIB buses. To generate RTN pulses, we use MATLAB's internal Poisson random number generator $poissrnd(\lambda)$ to obtain binary RTN sequences, and load the sequences into the Agilent 81150A AWG. Each binary RTN sequence consists of $50000$ data points and around $5000$ random jumps in average. The mean jumping rate $\chi$ (of AWG's output) is modulated by changing the clock frequency $\nu$ of the AWG: $\chi \approx 5000\times \nu$. To verify this relation between $\chi$ and $\nu$, we observe the RTN sequences at different $\nu$'s by a fast oscilloscope ($10$~GS/s), count the number $n$ of edges (jumping events) for certain period of time $t$, and calculate the real mean jumping rate by its original definition, $\chi = n/t$. As an example, a $0.5$~$\mu$s long RTN sequence with estimated mean jumping rate $\chi \approx 5000\times 10$~KHz~$= 50$~MHz is shown in Fig.~\ref{room_electronics}c. $25$ jumps are counted during $0.5$~$\mu$s, which gives $\chi = 50$~MHz. As long as $\nu <100$~KHz, the formula $\chi \approx 5000\times \nu$ gives a good estimation of the mean jumping rate. For $\nu >100$~KHz, the real mean jumping rate is smaller than the estimated one, due to the intrinsic $\sim 2$~ns rising/falling time of the AWG (pulses shorter than $4$~ns have a certain chance to merge together).

\section{Random modulation}
We consider the transmon qubit
   \begin{equation}
 \hat H(t) =\frac\hbar 2 [\omega_0+\Dev(t)]\hat{\sigma}_z , \label{eq:ham-time-dep}
\tag{1}
  \end{equation}
whose energy splitting has a fixed mean at $\omega_0/2\pi=2.62$  GHz and the time-dependent part $\hbar \xi(t)$ is modulated in as random telegraph noise (see the main text). We do not observe transitions to the higher transmon levels in the experiments with the current driving parameters (Fig.~2 of the main text) thus it is sufficient to include only the two lowest transmon eigenstates in the analytical and numerical analysis. A sample path of the time-dependent energy splitting is shown in Fig.~1c of the main text and in Fig.~\ref{room_electronics}c. The jumping between the two 'states' $\hbar(\omega_0\pm\xi)$ occurs at random times, that is, $\hbar\xi(t)$ is a stochastic process. We consider a special case of a stochastic process, namely a Poisson process, which is characterized by its mean jumping frequency $\chi$ (\#jumps/time) and by the probability $P_n(t)=(\chi t)^n \ee^{-\chi t}/n!$ for exactly $n$ jumps within a time interval. Another way to characterize the stochastic process is to construct the autocorrelation function $R(\tau)$,
\begin{align}
R(\tau)\equiv\langle \Dev(t+\tau)\Dev(t) \rangle_\Dev=\Dev^2\sum_{n=0}^{\infty}(-1)^n P_n(|\tau|)=\Dev^2\ee^{-\JumpR|\tau|}\sum_{n=0}^{\infty} \frac{(-\JumpR |\tau|)^n}{n!}=\Dev^2\ee^{-2\JumpR|\tau|},\label{R:alpha}
\end{align}
which defines the correlation time of the displacement process: $\tau_\Dev=1/2\JumpR$. As a notational convention, we will use the notation $\langle \Dev(t+\tau)\Dev(t) \rangle_\Dev$ to denote an average over the fluctuation process $\xi(t)$ without the risk of confusion with the usual ensemble average.

\subsection{The qubit spectrum}
The absorption spectrum of a qubit is defined as~\cite{GardinerZoller}
\begin{equation}
  S(\omega)=\frac{1}{2\pi}\int_{-\infty}^{\infty} \ee^{\ii \omega \tau}\langle\hat{\sigma}_-(\tau)\hat{\sigma}_+(0)\rangle_\Dev \D{\tau}. \label{spect}
\end{equation}
The correlation function $\langle\hat{\sigma}_-(\tau)\hat{\sigma}_+(0)\rangle_{\Dev}$  is calculated by tracing over the environmental degrees of freedom and averaging over the random telegraph noise (RTN) modulations. The former, denoted as $\langle\hat{\sigma}_-(\tau)\hat{\sigma}_+(0)\rangle$, can be evaluated by using the quantum regression theorem~\cite{GardinerZoller}, which states that if the equations of motions for the expectation values of a set of system operators are expressed as
\begin{equation}
  \frac{\D{}}{\D{t}}\langle\hat{A}_i(t)\rangle=\sum_j G_{ij}(t)\langle\hat{A}_j(t)\rangle, \label{eqn:mot}
\end{equation}
then for the correlations the following relations hold
\begin{equation}
  \frac{\D{}}{\D{t}}\langle\hat{A}_i(t+\tau)\hat{A}_k(t)\rangle=\sum_j G_{ij}(\tau)\langle\hat{A}_j(t+\tau)\hat{A}_k(t)\rangle, \label{eqn:mot:corr}
\end{equation}
where $\langle\hat{A}_i(t)\rangle=\textrm{Tr}\left\{\hat{\rho}(t)\hat{A}_i\right\}$ and $\hat{\rho}$ denotes the density matrix of the system. This means explicitly that if we know the equation of motion for $\langle\hat{\sigma}_-(\tau)\rangle$, then we know it for the correlation $\langle\hat{\sigma}_-(\tau)\hat{\sigma}_+(0)\rangle$ as well; only the initial value may differ. From the master equation
\begin{equation}
\frac{\D{\hat{\rho}(t)}}{\D{t}}=-\frac{\ii}{\hbar}\left[\hat{H}(t),\hat{\rho}(t)\right]+\frac{\Gamma_1}{2}\left(2\hat{\sigma}_-\hat{\rho}\hat{\sigma}_+-\hat{\sigma}_+\hat{\sigma}_-\hat{\rho}-\hat{\rho}\hat{\sigma}_+\hat{\sigma}_-\right) -\frac{\Gamma_\varphi}{4}[\hat{\sigma}_z,[\hat{\sigma}_z,\hat{\rho}]],
\end{equation}
we now first build the corresponding equation~\eqref{eqn:mot} for $\langle\hat{\sigma}_-(t)\rangle$:
\begin{equation}
\frac{\D{}}{\D{t}}\langle\hat{\sigma}_-(t)\rangle=\textrm{Tr}\left\{\frac{\D{\hat{\rho}(t)}}{\D{t}} \hat{\sigma}_-\right\}=\left(-\ii \omega_0-\ii \xi(t)-\Gamma_1-\frac{\Gamma_\phi}{2}\right)\langle\hat{\sigma}_-(t)\rangle.
\end{equation}
Then, by the quantum regression theorem applied to equations~\eqref{eqn:mot}-\eqref{eqn:mot:corr}, we can write the corresponding equation for the correlation function  $\langle\hat{\sigma}_-(\tau)\hat{\sigma}_+(0)\rangle$ and solve it,
\begin{equation}
\langle\hat{\sigma}_-(\tau)\hat{\sigma}_+(0)\rangle=\ee^{-\ii\omega_0\tau-\Gamma_2|\tau|} \ee^{-\ii \int_0^\tau \Dev(\tau)\D{t}},
\end{equation}
where we have set the initial value $\langle\hat{\sigma}_-(0)\hat{\sigma}_+(0)\rangle=1$. The average over the fluctuations is denoted formally as
\begin{equation}
\langle\hat{\sigma}_-(\tau)\hat{\sigma}_+(0)\rangle_\Dev=\ee^{-\ii\omega_0\tau-\Gamma_2|\tau|} \left\langle \ee^{-\ii \int_0^\tau \Dev(\tau)\D{t}}\right\rangle_\Dev,
\end{equation}
which gives the spectrum in equation~\eqref{spect}
\begin{align}
  S(\omega) &=\frac{1}{2\pi}\int_{-\infty}^{\infty}\ee^{\ii(\omega-\omega_0)\tau-\Gamma_2|\tau|} \left\langle \ee^{-\ii\int_0^\tau\Dev(t)\D{t}}\right\rangle_\Dev\D{\tau} \label{eq:spect-random} \tag{2}\\
&=\frac{1}{2\pi}\int_{-\infty}^{\infty}\ee^{\ii(\omega-\omega_0)\tau-\Gamma_2|\tau|} \left\langle f(\tau)\right\rangle_\Dev\D{\tau}.
  \end{align}

 The next step is to calculate the average over the RTN fluctuations in $ f(\tau)=\exp(-\ii \int_0^\tau \Dev(\tau)\D{t})$ to get the explicit expression for the spectrum~\cite{Anderson54, Abragam, Laser84}. We start by forming a differential equation $\dot{f}(\tau)=-\ii\Dev(\tau)f(\tau)$, which, after a single iteration, assumes the following form
\begin{equation}
\frac{\partial f(\tau)}{\partial \tau}=-\ii \Dev(\tau)-\int_0^\tau\Dev(\tau)\Dev(t)f(t)\D{t}.
\end{equation}
Now, by exploiting the symmetry of the jumping process $\langle\Dev(t)\rangle_\Dev=0$ and the factorization rule~\cite{Laser84}
\begin{equation}
\langle \Dev(t_1)\Dev(t_2)\ldots\Dev(t_n)\rangle_\Dev=\langle \Dev(t_1)\Dev(t_2)\rangle_\Dev\langle\Dev(t_3)\Dev(t_4)\ldots\Dev(t_n)\rangle_\Dev, \label{fact_rule}
\end{equation}
where $t_1<t_2<\ldots<t_n$, we write the differential equation for the average of the function $f(\tau)$
\begin{equation}
\frac{\partial \langle f(\tau) \rangle_\Dev }{\partial \tau}=-\int_0^\tau R(\tau-t)\langle f(t) \rangle_\Dev \D{t}.\label{integro-differential}
\end{equation}
Noteworthily, the displacement process $\hbar\Dev(t)$ does not obey Gaussian statistics in spite of the identical autocorrelations, implying that the usual Gaussian cumulant expansion cannot be applied~\cite{Laser84}. In other words, the above factorization rule in equation~\eqref{fact_rule} does not hold for a Gaussian displacement process~\cite{Laser84}, whose corresponding correlation functions of order higher than two ($n>2$) vanish. With an application of Laplace transformation to the integro-differential equation~\eqref{integro-differential}, we notice that the Laplace transform $\widetilde{F}(s)$ of  $\langle f(\tau) \rangle_\Dev$ obeys a simple algebraic rule
\begin{equation}
\widetilde{F}(s)=\frac{1}{s+\widetilde{R}(s)}, \label{lap.f}
\end{equation}
expressed with the Laplace transform $\widetilde{R}(s)$ of the autocorrelation function $R(\tau)$ (see Ref.~\onlinecite{Sagi10} for an alternative derivation).

Up to this point, the consideration has been general for any Markovian two-state noise process. The Fourier transform of time-symmetric functions  can be calculated in an elegant way by exploiting the identity $F(\omega)=\pi^{-1} \textrm{Re}[\widetilde{F}(s=-\ii\omega)]$. By using the explicit expression for the autocorrelation function $R(\tau)$ in equation~\eqref{R:alpha}, the Laplace transformation becomes $\widetilde{R}(s)=\Dev^2 (s+2\JumpR)^{-1}$. Thus, the Fourier transform of the function $\langle f(\tau) \rangle_\Dev$ is
\begin{equation}
F(\omega)=\frac{1}{\pi} \frac{2\JumpR\Dev^2}{(\omega^2-\Dev^2)^2+4\JumpR^2 \omega^2}. \label{F:motaveg}
\end{equation}
The explicit expression in time domain for the expectation value $\langle f(\tau) \rangle_\Dev$ is
\begin{equation}
\langle f(\tau) \rangle_\Dev=\ee^{-\JumpR|\tau|}\left[\frac{\JumpR}{\sqrt{\JumpR^2-\Dev^2}} \sinh \left(\sqrt{\JumpR^2-\Dev^2} |\tau|\right)+\cosh \left(\sqrt{\JumpR^2-\Dev^2} |\tau|\right)\right],
\end{equation}
which is obtained with the inverse Laplace (Fourier) transformation. Here, one clearly observes the change of the dynamics in $\langle f(\tau) \rangle_\Dev$ as the dynamical threshold, defined by $\JumpR=\Dev$, is crossed. The function $\langle f(\tau) \rangle_\Dev$ is oscillatory with jumping rates $\JumpR<\Dev$, but it becomes a decaying function when $\JumpR>\Dev$.

The spectrum in equation (3) of the main paper is then given by the convolution of $F(\omega)$ in equation~\eqref{F:motaveg} with a standard $\omega=\omega_0$ centered Lorentzian lineshape $S_{\omega_0,\Gamma_2}(\omega)$ with full width at half maximum $2\Gamma_2$,
\begin{align}
S(\omega)&=\frac{1}{2\pi}\int_{-\infty}^{\infty}\ee^{\ii(\omega-\omega_0)\tau-\Gamma_2|\tau|} \left\langle \ee^{-\ii\int_0^\tau\Dev(t)\D{t}}\right\rangle_\Dev\D{\tau} \tag{2}\label{S:conv:motaveg}\\ &=\int_{-\infty}^{\infty}S_{\omega_0,\Gamma_2}(\omega')  F(\omega-\omega')\D{\omega'}=\frac{1}{\pi}\frac{2\JumpR\Dev^2+\Gamma_2\left[(\Gamma_2+2\JumpR)^2+(\omega-\omega_0)^2+\Dev^2\right]}{(\Dev^2-(\omega-\omega_0)^2+\Gamma_2(\Gamma_2+2\JumpR))^2+4(\Gamma_2+\JumpR)^2(\omega-\omega_0)^2} \tag{3}. \label{S:spect_full}
\end{align}
We derive the asymptotic limits of slow and fast jumping by considering the convolution form in equation \eqref{S:spect_full}. By reducing $F(\omega)$ in equation \eqref{F:motaveg} at the limit $\chi\ll\xi$ around $\omega=\pm\Dev$ and at the limit $\chi\gg\xi$ around $\omega=0$, one gets that
\begin{align}
F(\omega)&=\frac{1}{2\pi} \frac{\JumpR}{(\omega\mp\Dev)^2+\JumpR^2}, & F(\omega)&=\frac{1}{\pi} \frac{\frac{\Dev^2}{2\JumpR}}{\omega^2+\left(\frac{\Dev^2}{2\JumpR}\right)^2},
\end{align}
respectively. Combining these with the Lorentzian lineshape $S_{\omega_0,\Gamma_2}(\omega')$ with the help of convolution in equation~\eqref{S:spect_full}, one obtains the total decoherence rates in the limits of slow and fast jumping $\Gamma_2'=\Gamma_2+\JumpR$ and $\Gamma_2'=\Gamma_2+\Dev^2/2\JumpR$, respectively. Note that our convention for the units of frequency is that the jumping rate $\chi$ is measured in frequency (Hz) and the other quantities, such as the jumping amplitude $\xi$ or the decoherence rate $\Gamma_2$, in angular frequency (rad/s).

\begin{figure}
\includegraphics[width=0.45\linewidth]{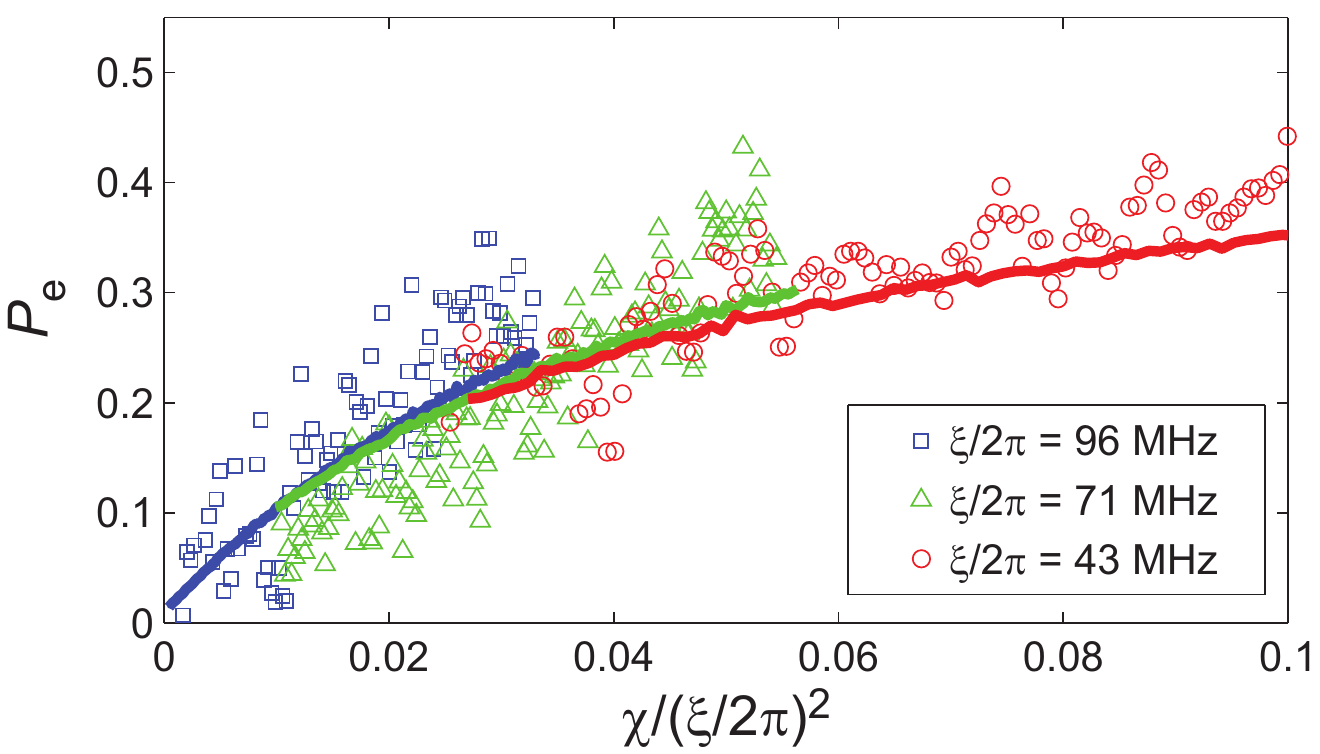}
\caption{\label{zero-detuning}{\bf Zero-detuning ($\omega = \omega_0$) occupation probability $P_{\rm e}$ versus jumping rate.} The horizontal axis is the scaled jumping rate $\chi / (\xi / 2\pi)^2$. The markers indicate measured $P_{\rm e}$ at $\xi/2\pi = 96$ MHz (blue squares), $\xi/2\pi = 71$ MHz (green triangles) and $\xi /2\pi = 43$ MHz (red circles). The solid lines are the corresponding simulated $P_{\rm e}$ with the same driving and decoherence parameters as in Fig.~2a of the main text.}
\end{figure}

In Fig.~\ref{zero-detuning} we present the measured zero-detuning ($\omega=\omega_0$) occupation probability $P_e$ at different measured jumping amplitudes $\xi$ together with the corresponding numerical simulations (see Methods of the main text). The data sets show similar gradually increasing behaviour, and by plotting them as a function of the scaled jumping rate $\chi/\xi^2$ they all fall onto same curve. This universal behaviour is explained by the absorption (without decoherence) in equation~\eqref{F:motaveg} at the zero detuning $F(\omega=0)=2\chi/\pi\xi^2$.  The absorption increases linearly in $\chi/\xi^2$ until qubit saturation (reaching the steady state occupation probability $P_e=0.5$ by the moderately strong drive $g>\Gamma_2$) for $\chi/(\xi/2\pi)^2\gtrsim0.03$~$(\textrm{MHz})^{-1}$.

\subsection{Master equation for a driven qubit}
When the qubit under RTN modulation of the energy splitting  is driven with a relatively strong transverse (Rabi) drive, $g>\Gamma_2$, equation~\eqref{S:spect_full} is not sufficient to describe the absorption since it does not include the saturation effects (power broadening) originating in the strong transverse driving. To overcome this limitation, we build up a master equation where the longitudinal RTN modulation is taken into account exactly in the qubit dynamics. We start by writing the total Hamiltonian (see equations (1) and (4) of the main paper) in the frame rotating with the driving frequency $\omega$ and assuming the rotating wave approximation (RWA),
\begin{equation}
  \hat{H}_{\rm tot}=\hat{H}_{\rm q}+\hat{H}_\Dev(t)+\hat{H}_r=\frac{\hbar}{2}\left[(\omega_0-\omega)\hat{\sigma}_z+g \hat{\sigma}_x\right]+\frac{\hbar}{2}\Dev(t)\hat{\sigma}_z+\hat{H}_r, \label{ham:reserv}
\end{equation}
where $\hat{H}_{\rm q}$ denotes the driven qubit, $\hat{H}_r$ is the Hamiltonian of the reservoir (environment) coupling to qubit via $\hat{H}_{\Dev}(t)$. Now, we assume that the displacement process $\hbar\Dev(t)$ is caused by coupling to the environment (bath). The intention is to trace the bath out and get equations of motion for the density matrix $\hat{\rho}$ of the qubit only.

In the interaction picture (denoted with prime), the equation of motion for the density matrix $\hat{w}$ of the total system reads
\begin{equation}
  \frac{\D{}}{\D{t}}\hat{w}'(t)=\frac{1}{\ii\hbar}[\hat{H}'_{\Dev}(t),\hat{w}'(t)]\label{master.prime}.
\end{equation}
By iterating~\eqref{master.prime} and taking the average over fluctuations in $\hat{H}_{\Dev}(t)$ (trace over the bath $\hat{H}_r$), the density matrix $\hat{\rho}'$ becomes
\begin{equation}
 \hat{\rho}'=\hat{\rho}'(0)+\sum_{n=1}^\infty \left(\frac{1}{\ii\hbar}\right)^n \int_0^t \D{\tau_1} \int_0^{\tau_1} \D{\tau_2}\cdots\int_0^{\tau_{n-1}}\D{\tau_n} \left\langle [\hat{H}'_{\Dev}(\tau_1),[\hat{H}'_{\Dev}(\tau_2),\ldots[\hat{H}'_{\Dev}(\tau_n),\hat{\rho}'(0)]]]\right\rangle_{\Dev}.
\end{equation}
Now, we adopt the notation $\hat{L}'_\Dev(t)=[\hat{H}'_\Dev(t),(\cdot) ]$ for the commutator, which can be also understood as a $3\times3$ matrix if the density operator $\hat{\rho}'$ is seen as a $3\times 1$ vector. By assuming the statistical properties described by the factorization rule in equation~\eqref{fact_rule}, the above iteration can be truncated~\cite{Mukamel78}, such that we obtain the master equation
\begin{equation}
  \frac{\D{}}{\D{t}}\hat{\rho}'(t)=-\frac{1}{\hbar^2}\int_0^t \left\langle\hat{L}'_\Dev(t)\hat{L}'_\Dev(\tau) \right\rangle_\Dev\hat{\rho}'(\tau)\D{\tau}.
\end{equation}
Transforming back to the Schr\"odinger picture results in
\begin{equation}
 \frac{\D{}}{\D{t}}\hat{\rho}(t)=\frac{1}{\ii\hbar}\hat{L}_q\hat{\rho}(t)-\frac{1}{\hbar^2}\int_0^t \left\langle\hat{L}_\Dev(t)\ee^{-\ii \hat{L}_q(t-\tau)}\hat{L}_\Dev(\tau) \right\rangle_\Dev\hat{\rho}(\tau)\D{\tau},
\end{equation}
where $\hat{L}_q\rho(t)=[\hat{H}_q, \rho(t)]$ and $\ee^{-\ii\hat{L}_q(t-\tau)}$ is the time evolution operator, that is, the formal solution of the master equation $\dot{\hat{\rho}}(t)=\hat{L}_q\hat{\rho}(t)/\ii\hbar$. Using now the vector representation for the density matrix $\vekt{\rho}=(\rho_z, \rho_{eg}, \rho_{ge})^T$, where $\rho_z=\rho_{ee}-\rho_{gg}$, we have the following matrices corresponding to the Hamiltonian in equation~\eqref{ham:reserv}
\begin{align*}
  \matr{L}_q&=\hbar \begin{pmatrix}  0 & -g & g \\  -\frac{g}{2} & \omega_0-\omega & 0 \\ \frac{g}{2} & 0 & -(\omega_0-\omega) \end{pmatrix}, & \matr{L}_\Dev(t)&=\hbar \begin{pmatrix} 0 & 0 & 0 \\ 0 & \Dev(t) & 0 \\ 0 & 0 & -\Dev(t) \end{pmatrix}.
\end{align*}
The subsequent equation needs only information on the autocorrelation $R(\tau)$ of the jumping process $\Dev(t)$ in equation~\eqref{R:alpha} and the vector-matrix algebra,
\begin{equation}
 \frac{\D{\vekt{\rho}(t)}}{\D{t}}=\frac{1}{\ii\hbar}\matr{L}_q\vekt{\rho}(t)-\frac{1}{\hbar^2}\int_0^t  \left\langle\matr{L}_\Dev(t)\ee^{-\ii \matr{L}_q(t-\tau)}\matr{L}_\Dev(\tau) \right\rangle_\Dev\vekt{\rho}(\tau)\D{\tau}.
\end{equation}
To calculate the matrix exponent $\exp[-\ii\matr{L}_q(t-\tau)]$, we do a transformation to the eigenbasis of $\matr{L}_q$, where the matrix exponent has only entries $1$ and $\exp[\mp\ii\sqrt{(\omega_0-\omega)^2+g^2}(t-\tau)]$ on its diagonal. Then we transform the master equation back to calculation basis, add the standard decoherence rates $\Gamma_{1,2}$, and express it using the Bloch pseudo-spin representation
\begin{align}
\dot{S}_x(t)&=[{\bf \Omega}\times{\bf S}(t)]_x-\int_0^t R(\tau)\left(\cos(g_{\rm d} \tau) S_x(t-\tau)+\frac{|\omega_0-\omega|}{g_{\rm d}}\sin(g_{\rm d} \tau) S_y(t-\tau)\right)\D{\tau}-\Gamma_2 S_x(t),\label{spin-master-x}\\
\dot{S}_y(t)&=[{\bf \Omega}\times{\bf S}(t)]_y-\int_0^t R(\tau)\left(\frac{g^2+(\omega-\omega_0)^2\cos(g_{\rm d} \tau)}{g_{\rm d}^2} S_y(t-\tau)-\frac{|\omega_0-\omega|}{g_{\rm d}}\sin(g_{\rm d} \tau) S_x(t-\tau)\right)\D{\tau} -\Gamma_2 S_y(t),\\
\dot{S}_z(t)&=[{\bf \Omega}\times{\bf S}(t)]_z-\Gamma_1 (S_z(t)-S_{z0}), \label{spin-master-z}
\end{align}
where ${\bf \Omega}=(g,0,\omega_0-\omega)$ and $g_{\rm d}=\sqrt{(\omega_0-\omega)^2+g^2}$ denotes the detuned Rabi-frequency.

In general, the correlation time $\tau_\Dev=1/2\JumpR$ of the displacement process $\Dev(t)$ can have values comparable to the system dynamics (Rabi frequency $g$ is of the order of $2\pi\times 10$ MHz), since in our measurements the jumping rate $\JumpR$ varies between $50$ MHz and $600$ MHz. In general, the Markov approximation [short correlations in $R(\tau)$] is not valid for solving equations~\eqref{spin-master-x}-\eqref{spin-master-z}. The Markov approximation can be used only, when the autocorrelation $R(\tau)$ has non-negligible values near $\tau=0$. This is the fast-jumping case, $\chi\gg\xi$, in which situation the Bloch equations become
\begin{align}
\dot{S}_x(t)&=[{\bf \Omega}\times{\bf S}(t)]_x-\left(\Gamma_2+\frac{\Dev^2}{2\JumpR}\right) S_x(t), \label{RTN-bloch1}\\
\dot{S}_y(t)&=[{\bf \Omega}\times{\bf S}(t)]_y-\left(\Gamma_2+ \frac{\Dev^2}{2\JumpR}\right) S_y(t),\\
\dot{S}_z(t)&=[{\bf \Omega}\times{\bf S}(t)]_z-\Gamma_1 (S_z(t)-S_{z0}), \label{RTN-bloch3}
\end{align}
describing the Rabi-drive with the detuning $\omega_0-\omega$, the driving amplitude $g$, the relaxation rate $\Gamma_1$, and the total decoherence rate $\Gamma_{2}'=\Gamma_2+\Dev^2/2\JumpR$. The time-dependent solution of equations~\eqref{RTN-bloch1}-\eqref{RTN-bloch3} can be used to reproduce the measured Rabi oscillations (Fig.~4a of the main text). The numerical simulation (shown in Fig.~4a) includes the imperfections of the experimental wave forms (see Methods of the main text), giving very good quantitative agreement with the data.

\section{Sinusoidal modulation}

The effects of sinusoidally modulating the transition frequency of an atom is a generic problem in theoretical physics, and has attracted a lot of interest in the past, most notably in the quantum-optics community \cite{Nayak85, Kador00}. We address now this problem in the context of superconducting qubits. We consider a transmon qubit whose energy splitting is modulated sinusoidally: $\hbar[\omega_0+\delta\cos(\Omega t)]$. Even though $\Omega\ll \omega_0$, the qubit can be excited with the low-frequency signal in the presence of an additional high-frequency drive $\omega$. This can be seen as a multi-photon transition process involving quanta from both fields [in equation~\eqref{eq:mod} of the main text], or, as a photon-assisted  Landau-Zener-St\"uckelberg-interference~\cite{Oliver05, Sillanpaa06, Nori10} (LZS) between the dressed states $|\downarrow,n\rangle$ and $|\uparrow,n-1\rangle$.

\subsection{Multi-photon transition processes}
\begin{figure}
  \includegraphics[width=0.7\linewidth]{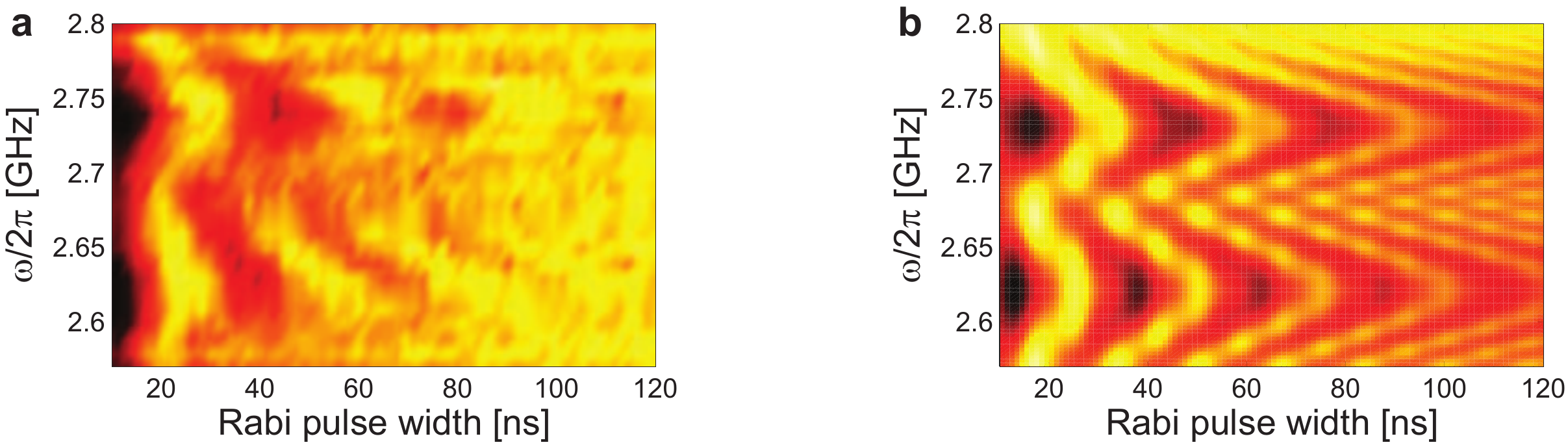}
\caption{\label{rabi_sin}{\bf Rabi oscillations with sinusoidal modulation.} {\bf a,} Experimental data for the $0$-band (centered at $\omega_0/2\pi = 2.62$~GHz) and for the $1$-band (centered at $\omega_0 + \Omega$) taken with modulation amplitude $\delta/2\pi \approx 140$~MHz and modulation frequency $\Omega/2\pi = 120$~MHz. {\bf b,} Numerical simulation. The Rabi frequency without modulation is $g/2\pi = 60$~MHz. The effective Rabi frequency (with detuning) for the $k$-band is $g_{\rm eff}^{(k=0,1)} = \sqrt{(\omega_0 + k\Omega - \omega)^2 + g^2 J_k^2(\delta/\Omega)}$ in equation~\eqref{rabi:sin:mod}. The decoherence parameters used for the numerical calculation are $\Gamma_1/2\pi = 1$~MHz and $\Gamma_2/2\pi = 3$~MHz. }
\end{figure}
Let us first consider the multi-photon transition processes associated with the sinusoidal-modulated Hamiltonian, defined in equations (1) and (4) of the main text
\begin{equation}
  \hat{H}=\frac{\hbar}{2}[\omega_0+\delta \cos (\Omega t)]\hat{\sigma}_z+\hbar g \cos (\omega t) \hat{\sigma}_x. \label{sin:ham}
\end{equation}
The qubit is driven both in the longitudinal ($\hat{\sigma}_z$) and in the transverse ($\hat{\sigma}_x$) direction. We show below how to eliminate the time-dependence from the longitudinal drive. After moving to a non-uniformly rotating frame with the unitary transformation~\cite{Oliver05,Tuorila10}
\begin{equation}
  \hat{U}=\exp\left[-\frac\ii2  \left(\omega_0 t+\frac\delta\Omega \sin\Omega t\right)\hat{\sigma}_z\right],
\end{equation}
the effective Hamiltonian is $\hat{H}'=\hat{U}^\dagger \hat{H}\hat{U}+\ii\hbar(\partial_t {\hat{U}}^\dagger) \hat{U}$, and by using the Jacobi-Anger relations, we get
\begin{equation}
  \hat{H}'=\frac{\hbar g}{2} \left(\ee^{\ii\omega t}+\ee^{-\ii\omega t}\right)\ee^{\ii\omega_0 t}\sum_{k=-\infty}^\infty J_k\left(\frac \delta \Omega \right) \ee^{\ii k \Omega t} \hat{\sigma}_++ \textrm{h.c.}.
\end{equation}
By assuming that the transverse drive is close to a resonance, that is, $\omega\approx\omega_0\pm k\Omega$ ($k=0,1,2\ldots$) and that the resonances are resolvable $\Omega>g>\Gamma_2$,  we transform back with $\hat{U}=\exp[\ii(\omega_0+k\Omega-\omega)t\hat{\sigma}_z/2]$ and ignore all but the resonant terms, {\it i.e.}, all the fast rotating terms (rotating wave approximation, RWA). The resulting RWA Hamiltonian reads
\begin{equation}
  \hat{H}'_{\rm RWA}=\frac\hbar 2\left[(\omega_0+k\Omega-\omega)\hat{\sigma}_z+ g J_k\left(\frac \delta \Omega \right) \hat{\sigma}_x\right],
\end{equation}
describing, in the Bloch spin representation, precessions around the vector ${\bm \Omega}=\left(g J_k\left(\frac \delta \Omega \right), 0, \omega_0+k\Omega-\omega\right)$ with the effective Rabi frequency
\begin{equation}
g_{\rm eff}^{(k)}=\sqrt{(\omega_0-\omega+k\Omega)^2+g^2J^2_k\left(\delta/\Omega\right)}. \label{rabi:sin:mod}
\end{equation}
Exactly at the multi-photon resonance $\omega=\omega_0\pm k\Omega$ ($k=0,1,2\ldots$),  the effective Rabi frequency is $|g J_k(\delta/\Omega)|$, and it is plotted in Fig.~4c of the main text.
\begin{figure}
  \includegraphics[width=0.8\linewidth]{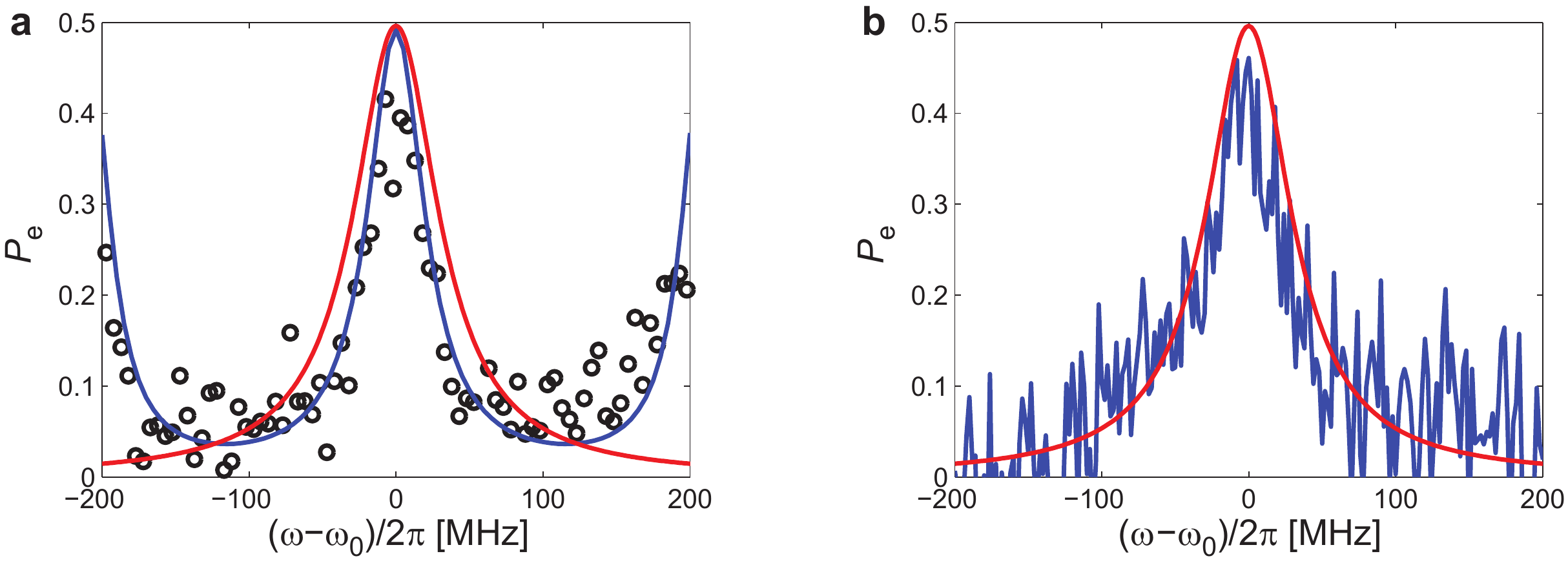}
\caption{\label{spectro_compare} {\bf Excited-state occupation probabilities with and without modulation.} {\bf a,} The same as Fig.~4c in the main text, except for the red curve that indicates occupation probability without any modulation in equation~\eqref{eq:non-mod}, with simulation parameters $g/2\pi = 20$~MHz, $\Gamma_1/2\pi = 1$~MHz, and $\Gamma_2/2\pi = 3$~MHz. {\bf b,} The blue curve denotes the measured occupation probability without any modulation and with the same driving amplitude as that in {\bf a}; the red curve is the the same as the red curve in {\bf a}. The linewidth of the $0$-band (blue curve in {\bf a}), $\sqrt{\Gamma_2^2 + \Gamma_2\left[ g J_0\left( \delta/\Omega \right) \right]^2/\Gamma_1}$ [see equation~\eqref{eq:mod}], is always smaller than the linewidth of the bare qubit (red curves in {\bf a,b}), $\sqrt{\Gamma_2^2 + \Gamma_2 g^2/\Gamma_1}$ [see equation~\eqref{eq:non-mod}], since the Bessel function $J_0(\delta/\Omega)<1$  except at $\delta = 0$.}
\end{figure}

In the time domain, the measured Rabi oscillations are presented in Fig.~\ref{rabi_sin} together with numerical simulations of Bloch equations exploiting the effective Rabi frequency in equation~\eqref{rabi:sin:mod}. To find the steady state occupation probability in the presence of relaxation with rate $\Gamma_1$ and decoherence with rate $\Gamma_2$, we solve the Bloch equations analytically.  In the RWA, that is when $\Omega>g>\Gamma_2$ is satisfied, one is allowed to add up independent contributions from all the resonances (the resolved sidebands). The result for the steady state excited state occupation probability is
\begin{equation}
  P_{\rm e}= \sum_{k=-\infty}^\infty\frac{\frac{\Gamma_2}{2\Gamma_1} \left[g J_k\left(\frac{\delta}{\Omega}\right)\right]^2}{\Gamma^2_2+(\omega_0-\omega+k\Omega)^2+\frac{\Gamma_2}{\Gamma_1}\left[g J_k\left(\frac{\delta}{\Omega}\right)\right]^2} \tag{5}. \label{eq:mod}
\end{equation}
In the non-modulated case, the corresponding expression for the occupations is simply
\begin{equation}
  \label{eq:non-mod}
  P_{\rm e}=\frac{\frac{\Gamma_2}{2\Gamma_1}g^2}{\Gamma_2^2+(\omega_0-\omega)^2+\frac{\Gamma_2}{\Gamma_1} g^2}.
\end{equation}
The occupation probabilities in equation~\eqref{eq:mod} and~\eqref{eq:non-mod} are compared with the experimental steady state occupation probabilities in Fig.~\ref{spectro_compare}. An interesting effect (Fig.~\ref{spectro_compare}b) is that the linewidth of the modulated qubit on the central band is always smaller than the linewidth on the qubit in the absence of the modulation due to reduced power broadening.

\subsection{Photon-assisted Landau-Zener-St\"uckelberg interference}

\begin{figure}
\includegraphics[width=0.95\linewidth]{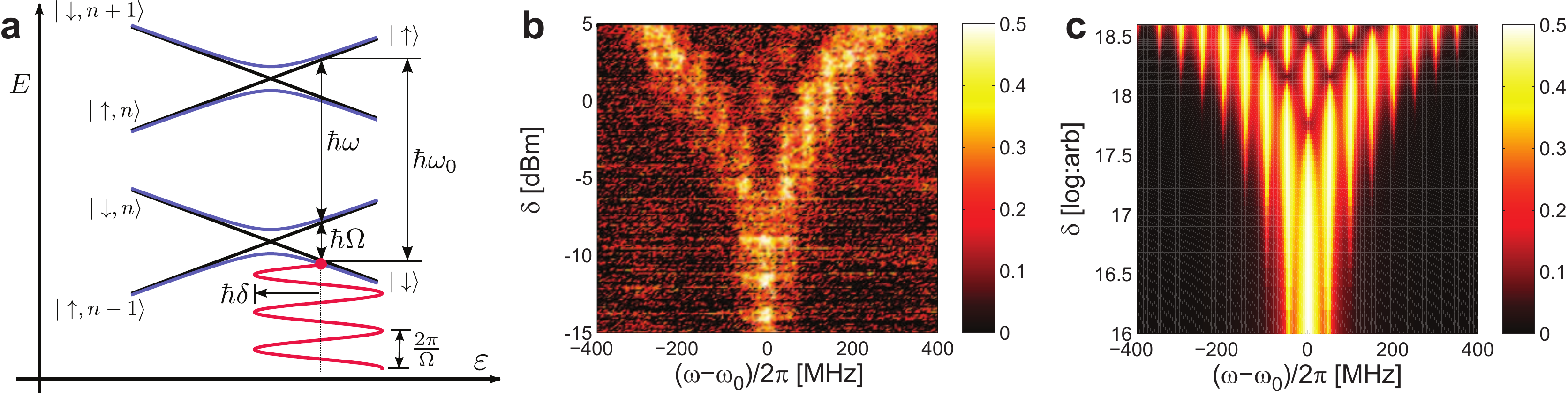}
\caption{\label{LZSfig} {\bf Photon-assisted Landau-Zener-St\"uckelberg interference.} {\bf a,} Schematics of the photon-assisted LZS interference. The black diagonal lines denote the transversely dressed energy levels of the transmon $\left |\uparrow/\downarrow, n\right\rangle$, shifted with photon quanta $\pm n\hbar\omega$ from the original energy levels of the transmon. The red curve shows the $2\pi/\Omega$-periodic time-evolution of the energy splitting $\varepsilon=\hbar[\omega_0+\delta\cos(\Omega t)]$. The blue curves are the adiabatic energy levels of the Hamiltonian in equation~\eqref{LZS-ham}, which the system would follow in the absence of LZS-tunneling. {\bf b,} Measured excited-state occupation probability as a function of the sinusoidal wave amplitude $\delta$ (in~dBm) added on top of dc flux bias, with fixed modulation frequency $\Omega/2\pi = 50$~MHz. {\bf c,} Simulation corresponding to panel {\bf b}, with parameters $\Gamma_1/2\pi=1$ MHz, $\Gamma_2/2\pi=3$ MHz, and $g/2\pi=20$ MHz. }
\end{figure}

The spectra seen in the experiment under sinusoidal modulation can be also interpreted as a photon-assisted Landau-Zener-St\"uckelberg (LZS) effect. Note that in the absence of the driving field standard LZS processes are not possible: indeed the qubit energy separation is one order of magnitude higher than the modulation frequency, therefore the standard LZS probability is negligibly small. However, the system can still perform LZS-transitions by absorbing a photon from the driving field. This photon-assisted LZS-interference can be seen by transforming the Hamiltonian $\hat{H}+\hat{H}_{\rm drive}$ from the equation~\eqref{sin:ham} into the frame rotating at $\omega$ around the $z$-axis [unitary transformation $\hat{U}=\exp(-\ii\omega \hat{\sigma}_z/2)$], where it has exactly the same form (in the RWA) as that of LZS-interference~\cite{Oliver05, Sillanpaa06, Nori10}, namely
\begin{equation}
\hat{H}=\frac{\hbar}{2}[\omega_0-\omega+\delta \cos (\Omega t)]\hat{\sigma}_z+\frac{\hbar}{2}g \hat{\sigma}_x. \label{LZS-ham}
\end{equation}
As illustrated in Fig.~\ref{LZSfig}a, LZS-transition events may occur when the modulation amplitude $\delta$ is of the order of, or larger than, the detuning between the driving frequency and the qubit splitting $\omega-\omega_0$. The LZS-transitions occur between the transversely dressed states $|\downarrow,n\rangle$ and $|\uparrow,n+1\rangle$, where $n$ refers to the photon number of the transverse driving field. The phase difference of the two states gathered between the consecutive tunneling events leads to either constructive or destructive interference observed as maxima or minima in the occupation probability of the excited state, shown in Fig.~\ref{LZSfig}b,c. The LZS-interference seen in our system
corresponds to the so called fast passage limit~\cite{Nori10}.  In this limit, the expression for the occupation probability can be calculated analytically~\cite{Nori10} and the result agrees exactly with our equation~\eqref{eq:mod}.

To prove experimentally that this picture is valid, we have scanned the modulation amplitude $\delta$ and the driving frequency $\omega$ at fixed $\Omega$.  We observe an interference pattern of the steady state occupation probability in good agreement with the theoretical prediction, see Fig.~\ref{LZSfig}b,c.

\subsection{Simulated ultrastrong coupling}
The Hamiltonian of the transmon qubit and the $\lambda/4$-resonator in the eigenbasis of the transmon~\cite{Koch07} is
\begin{equation}
\hat{H}=\frac{\hbar}{2}[\omega_0+\delta\cos(\Omega t)]\hat{\sigma}_z+\hbar g \cos(\omega t) \hat{\sigma}_x+\hbar\omega_{\rm r} \hat{a}^\dagger\hat{a}+\hbar \kappa (\hat{\sigma}_-\hat{a}^\dagger+\hat{\sigma}_+ \hat{a}). \label{ham.total}
\end{equation}
 Here the third term is the Hamiltonian of our coplanar waveguide resonator with frequency $\omega_{\rm r}/2\pi=3.795$ GHz, while the fourth term is the qubit-resonator coupling with the strength $\kappa$. From single-tone qubit-resonator spectroscopy, we
can extract a value $\kappa/2\pi\approx 80$ MHz. This Hamiltonian assumes also a RWA in the coupling term. The transmon-resonator system is in the strong coupling regime and the Hamiltonian~\eqref{ham.total} as such describes the Jaynes-Cummings physics.  Our aim is that, by using suitably-chosen transformations and approximations, to show that the system~\eqref{ham.total} can be employed to simulate ultrastrong coupling, {\it i.e.} the regime where $\kappa/\widetilde{\omega}_{\rm r}$ is no longer negligible \cite{Solano10}. Here $\widetilde{\omega}_{\rm r}$ denotes the effective resonance frequency of the resonator, to be defined precisely below. The main idea is to take advantage of the two drivings of the qubit. The derivation follows the recent work by Ballester \textit{et~al.}~\cite{Ballester11}, with the difference that in our setup the drivings are in the longitudinal ($\hat{\sigma}_z$) and in the transverse ($\hat{\sigma}_x$) directions, while in the
 work mentioned above ~\cite{Ballester11} the authors have considered both drivings as transverse.

We start by making a transformation to the rotating frame with $\hat{U}=\exp\left(-\ii \omega\hat{\sigma}_z t/2\right)\exp\left(-\ii\omega \hat{a}^\dagger\hat{a}\right)$  resulting in the effective Hamiltonian $\hat{H}'=\hat{U}^\dagger\hat{H}\hat{U}+\ii\hbar(\partial_t \hat{U}^\dagger)\hat{U}$:
\begin{equation}
  \label{eq:ham_rw}
  \hat{H}'=\frac{\hbar}{2}[\omega_0-\omega+\delta \cos(\Omega t)]\hat{\sigma}_z+\frac{\hbar g}{2}\hat{\sigma}_x+\hbar \kappa (\hat{\sigma}_-\hat{a}^\dagger+\hat{\sigma}_+ \hat{a})+\hbar(\omega_{\rm r}-\omega)\hat{a}^\dagger\hat{a},
\end{equation}
where we have performed a RWA  in the transverse driving term. The next step is to move to the interaction picture with respect  to $\hat{H}^{g}=\frac{\hbar g}{2}\hat{\sigma}_x$, such that $\hat{H}^I=\ee^{\ii g t \hat{\sigma}_x/2} (\hat{H}-\hat{H}^g)\ee^{-\ii g t \hat{\sigma}_x/2}$. One needs the following auxiliary results:
\begin{align}
  \ee^{\ii g t \hat{\sigma}_x/2}\hat{\sigma}_z\ee^{-\ii g t \hat{\sigma}_x/2}&=\cos(g t) \hat{\sigma}_z+\sin(gt)\hat{\sigma}_y,\\ \ee^{\ii g t \hat{\sigma}_x/2}\hat{\sigma}_-\ee^{-\ii g t \hat{\sigma}_x/2}&=\frac{\ii}{2}\sin(gt)\hat{\sigma}_z+\cos^2\left(\frac{g t}{2}\right)\hat{\sigma}_-+\sin^2\left(\frac{g t}{2}\right)\hat{\sigma}_+,\\
\ee^{\ii g t \hat{\sigma}_x/2}\hat{\sigma}_x\ee^{-\ii g t \hat{\sigma}_x/2}&=\hat{\sigma}_x, \\ \ee^{\ii g t \hat{\sigma}_x/2}\hat{\sigma}_+\ee^{-\ii g t \hat{\sigma}_x/2}&=-\frac{\ii}{2}\sin(g t)\hat{\sigma}_z+\cos^2\left(\frac{g t}{2}\right)\hat{\sigma}_++\sin^2\left(\frac{g t}{2}\right)\hat{\sigma}_-.
\end{align}
In the interaction picture, we therefore get
\begin{align}
  \label{eq:ham_I}
  \hat{H}^I=&\frac{\hbar(\omega_0-\omega)}{2}\left[\cos(gt) \hat{\sigma}_z+\sin(gt)\hat{\sigma}_y\right]+\frac{\hbar\delta\cos(\Omega t)}{2}\left[\cos(gt) \hat{\sigma}_z+\sin(gt)\hat{\sigma}_y\right]+\hbar(\omega_{\rm r}-\omega)\hat{a}^\dagger\hat{a}
\\
&+\hbar \kappa \left(\left[\frac{\ii}{2}\sin(gt)\hat{\sigma}_z+\cos^2\left(\frac{gt}{2}\right)\hat{\sigma}_-+\sin^2\left(\frac{gt}{2}\right)
\hat{\sigma}_+\right]\hat{a}^\dagger+\left[-\frac{\ii}{2}\sin(gt)\hat{\sigma}_z+\cos^2\left(\frac{gt}{2}\right)\hat{\sigma}_++
\sin^2\left(\frac{gt}{2}\right)\hat{\sigma}_-\right]\hat{a}\right).\notag 
\end{align}
We now make a change of basis to $|\pm\rangle=(|g\rangle\pm|e\rangle)/\sqrt{2}$ using a  Hadamard transform
$\hat{\cal H}=\frac{1}{\sqrt{2}}\begin{pmatrix} 1 & 1\\ 1 &-1 \end{pmatrix}$, $\hat{H}^I_S=\hat{\cal H}\hat{H}\hat{\cal H}$:
\begin{align}
  \hat{H}^I_S=\frac{\hbar(\omega_0-\omega)}{2} \begin{pmatrix} 0 & \ee^{\ii gt} \\ \ee^{-\ii g t} & 0 \end{pmatrix}&+\frac{\hbar\delta}{4} \begin{pmatrix} 0 & \ee^{\ii(g+\Omega) t}+\ee^{\ii(g-\Omega) t} \\ \ee^{-\ii (g-\Omega) t}+\ee^{-\ii (g+\Omega) t} & 0 \end{pmatrix} \nonumber\\
&+\frac{\hbar\kappa}{2} \left[\begin{pmatrix} 1 &  \ee^{\ii g t} \\  -\ee^{-\ii g t} & -1 \end{pmatrix} \hat{a}^\dagger + \begin{pmatrix} 1 &  -\ee^{\ii g t} \\  \ee^{-\ii g t} & -1 \end{pmatrix}\hat{a} \right]+\hbar (\omega_{\rm r}-\omega)\hat{a}^\dagger \hat{a}. \label{S_frame}
\end{align}

The next important step is to choose $\Omega=g$ and make again a RWA by ignoring the time-dependent terms. This approximation becomes more and more
precise for larger values of the longitudinal driving frequency.
The final result is stunningly simple
\begin{equation}
\hat{H}^I_S=\frac{\hbar\delta}{4} \hat{\sigma}_x+\frac{\hbar \kappa}{2} \hat{\sigma}_z (\hat{a}+\hat{a}^\dagger)+\hbar (\omega_{\rm r}-\omega)\hat{a}^\dagger\hat{a},
\end{equation}
which, in the original $\{|g\rangle ,|e\rangle\}$ basis, reads
\begin{align}
\hat{H}^I&=\frac{\hbar\delta}{4} \hat{\sigma}_z+\frac{\hbar \kappa}{2} \hat{\sigma}_x (\hat{a}+\hat{a}^\dagger)+\hbar (\omega_{\rm r}-\omega)\hat{a}^\dagger\hat{a}=\frac{\hbar\widetilde{\omega}_0}{2} \hat{\sigma}_z+\hbar \widetilde{\kappa} \hat{\sigma}_x (\hat{a}+\hat{a}^\dagger)+\hbar \widetilde{\omega}_{\rm r}\hat{a}^\dagger\hat{a}. \label{S_Dicke}
\end{align}
One can obtain as well the result Eq. (\ref{S_Dicke}) directly from Eq. (\ref{eq:ham_I}) by neglecting the rotating terms under
the condition $\Omega=g$; however the change of basis resulting in Eq. (\ref{S_frame}) provides a motivation for this condition.

We have thus obtained, in Eq. (\ref{S_Dicke}), the standard single-qubit Dicke model of quantum optics (or the Jaynes-Cummings model without the RWA in the coupling term). The condition $\Omega =g$ is realized easily in our experiment, as the frequency 
of  the longitudinal modulation $\Omega/2\pi$ is swept over values ranging from 0 MHz to 500 MHz, while $g$ is the amplitude of the field that excites transversely the qubit, whose value can be calibrated from the Rabi frequency measurements. In this case, our system is mathematically equivalent to a qubit with Larmor frequency $\widetilde{\omega}_0 = \delta/2$, coupled by a coupling constant $\widetilde{\kappa} = \kappa /2$ to a harmonic oscillator with frequency $\widetilde{\omega}_{\rm r}=\omega_{\rm r}-\omega$. The energy splitting of the effective qubit $\hbar\widetilde{\omega}_0=\hbar\delta/2$ can be tuned through the amplitude of the longitudinal driving $\delta$. The simulated coupling ratio $\widetilde{\kappa}/\widetilde{\omega}_{\rm r}=\kappa/2(\omega_{\rm r}-\omega)$ can be tuned by changing the frequency of the transverse driving $\omega$ and it can reach values $\gtrsim 0.1$, that is, for these values the simulator enters the ultrastrong regime.
In our experiments, we have biased and operated the qubit as close as 200 MHz from the resonator, in other words $(\omega_{\rm r}-\omega)/2\pi=200$ MHz, resulting in an estimated effective $\widetilde{\kappa}/\widetilde{\omega}_{\rm r}\approx 0.2$ for the simulated coupling ratio. The fact the Larmor frequency of the effective qubit and the frequency of the harmonic oscillator can be easily and fast tuned experimentally by changing the amplitude of the longitudinal driving and, respectively, the frequency of the transverse driving of the real qubit opens up the realistic possibility of using this system as an analog simulator of quantum dynamics in the ultrastrong regime.